\documentclass[12pt,a4paper]{article}
\usepackage{amssymb}
\usepackage{epic}
\usepackage{eepic}
\def\bea{\begin{eqnarray}}
\def\eea{\end{eqnarray}}

\def\beq{\begin{equation}}
\def\eeq{\end{equation}}
\def\ba{\beq\new\begin{array}{c}}
\def\ea{\end{array}\eeq}
\def\be{\ba}
\def\ee{\ea}
\newcommand {\so}{\scriptscriptstyle{0}}
\newcommand {\spm}{\scriptscriptstyle{\pm}}
\newcommand {\spp}{\scriptscriptstyle{+}}
\newcommand {\sqp}{\scriptscriptstyle{1}}
\newcommand {\sqm}{\scriptscriptstyle{-1}}
\newcommand {\skv}{\scriptscriptstyle{2}}
\newcommand {\smkv}{\scriptscriptstyle{-2}}
\newcommand {\smf}{\scriptscriptstyle{-4}}
\newcommand {\spf}{\scriptscriptstyle{4}}

\makeatletter
\newdimen\normalarrayskip % skip between lines
\newdimen\minarrayskip % minimal skip between lines
\normalarrayskip\baselineskip
\minarrayskip\jot
\newif\ifold \oldtrue \def\new{\oldfalse}
\def\arraymode{\ifold\relax\else\displaystyle\fi} % mode of array entries
\def\eqnumphantom{\phantom{(\theequation)}} % right phantom in eqnarray
\def\@arrayskip{\ifold\baselineskip\z@\lineskip\z@
\else
\baselineskip\minarrayskip\lineskip2\minarrayskip\fi}
\def\@arrayclassz{\ifcase \@lastchclass \@acolampacol \or
\@ampacol \or \or \or \@addamp \or
\@acolampacol \or \@firstampfalse \@acol \fi
\edef\@preamble{\@preamble
\ifcase \@chnum
\hfil$\relax\arraymode\@sharp$\hfil
\or $\relax\arraymode\@sharp$\hfil
\or \hfil$\relax\arraymode\@sharp$\fi}}
\def\@array[#1]#2{\setbox\@arstrutbox=\hbox{\vrule
height\arraystretch \ht\strutbox
depth\arraystretch \dp\strutbox
width\z@}\@mkpream{#2}\edef\@preamble{\halign
\noexpand\@halignto
\bgroup \tabskip\z@ \@arstrut \@preamble \tabskip\z@ \cr}%
\let\@startpbox\@@startpbox \let\@endpbox\@@endpbox
\if #1t\vtop \else \if#1b\vbox \else \vcenter \fi\fi
\bgroup \let\par\relax
\let\@sharp##\let\protect\relax
\@arrayskip\@preamble}
\def\eqnarray{\stepcounter{equation}%
\let\@currentlabel=\theequation
\global\@eqnswtrue
\global\@eqcnt\z@
\tabskip\@centering
\let\\=\@eqncr
$$%
\halign to \displaywidth\bgroup
\eqnumphantom\@eqnsel\hskip\@centering
$\displaystyle \tabskip\z@ {##}$%
\global\@eqcnt\@ne \hskip 2\arraycolsep
$\displaystyle\arraymode{##}$\hfil
\global\@eqcnt\tw@ \hskip 2\arraycolsep
$\displaystyle\tabskip\z@{##}$\hfil
\tabskip\@centering
&{##}\tabskip\z@\cr}
\begingroup\ifx\undefined\newsymbol \else\def\input#1 {\endgroup}\fi

\begin{document}
\setcounter{footnote}{1}
\def\thefootnote{\fnsymbol{footnote}}
\begin{center}
\hfill ITEP-TH-38/01\\
\hfill hep-th/0110148\\
\vspace{0.3in}
{\Large\bf On Vertex Operator Construction of Quantum}

\vspace{0.1in}

{\Large\bf Affine Algebras
}
\end{center}
\bigskip
\bigskip
\centerline{{\large S. Klevtsov}
\footnote
{{e-mail: klevtsov@itep.ru}
}}

\bigskip

\centerline{\it Institute\,\,of\,\,Theoretical\,\,and\,\,Experimental\,\,
Physics,\,\,117259,\,\,Moscow,\,\,Russia}

\centerline{\it Department\,\,of\,\,Physics,\,\,Moscow\,\,State\,\,University,
\,\,119899,\,\,Moscow,\,\,Russia}
\bigskip
\abstract{\footnotesize
}
We describe a construction of the quantum deformed affine Lie algebras
using vertex operators in the free field theory. We prove Serre
relations for the quantum
deformed Borel subalgebras of affine algebras, namely
$ \widehat{\it sl}_{2}$ case is considered in detail. We provide some formulas
for the generators of affine algebra.


\begin{center}
\rule{5cm}{1pt}
\end{center}

\bigskip

\section {Introduction}

Vertex operator algebras \cite{Frenkel:1989im,Goddard,Halpern}
in the free field theories
\cite{Dotsenko,gerasimov,Felder:1989im} possess a rich mathematical structure.
Their connection with the representation theory of affine Lie algebras has been found
in \cite{ff} and with the theory of quantum groups in \cite{kanie}.
Various aspects of the
vertex operator realization of quantum groups were studied in
\cite{Bouwknegt:1990im,review}.

In this paper we study vertex operator construction of the quantum deformed
Borel subalgebras of
finite-dimensional and affine algebras in the theory of the free scalar
fields $\varphi^{i}, i=1,\ldots,n$. Free field theory is described by the
action
\be
\label{action}
S=\int d^{2}z\,(\partial\varphi^{i}\overline{\partial}\varphi^{i}+
i\alpha_{\so}R\rho^{i}\varphi^{i}),
\ee
where $R$ is a two-dimensional curvature for a background metric,
$\alpha_{\so}$
is a background charge and $\rho^{i}$
is some constant vector. We assume that $\varphi^{i}$ take values
in some appropriate torus. The root lattice associated with the torus gives rise to
a Lie algebra $g$ \cite{Goddard}. We consider vertex operators of
the conformal dimension one and investigate their commutation relations. They form an algebra
which is the quantum deformation of the original one.
Strictly speaking, if the underlying algebra has a Cartan decomposition
$g=n_{-}\oplus h \oplus n_{+}$ then the vertex operators corresponding to the
simple roots satisfy the identities of $U_{q}(n_{+})$ with some deformation parameter
$q=q_{\spp}$. In the same way the $U_{q}(n_{-})$ quantum group
(with another deformation parameter $q_{\scriptscriptstyle{-}}$)
can be obtained.

This type of construction for finite dimensional simply-laced algebras was
suggested in \cite{Felder:1989im} and studied
in \cite{Bouwknegt:1990im} (see \cite{review} for a review).
The aim of this work is to generalize the construction for
the case of affine quantum algebra. We proof the Serre relations and give explicit formulas for
generators in affine case.

Serre relations in quantum affine algebras have been studied also in
\cite{vs,varchenko}.

The vertex operators in the theory (\ref{action}) has a natural
classical limit $\alpha_{\so} \rightarrow 0$,
which should led to the classical vertex operators of \cite{Goddard}.
In this paper we use the approach in which classical
limit $\alpha_{\so} \rightarrow 0$
can be directly obtained (compare with \cite{Bouwknegt:1990im,review,feigin},
where it is not so explicit).

There is a well known close connection of the conformal minimal models and
quantum conformal Toda theory at the special values of the parameters.
One may hope that the construction proposed may be helpful for the
understanding of the quantization of conformal affine Toda theory
\cite{toda}.

The paper is organized as follows. In section 2 we describe
the construction of
$U_{q}( {\it sl}_{3})$. In section 3 we prove the Serre relations in this case.
The construction of the quantum deformed
affine algebra $U_{q}( \widehat{\it sl}_{2})$ is presented in sec. 4; details
of the calculation of Serre relations
and formulas for the generators in the affine case are collected in the appendix.
Let us note that the proof of the Serre relations for $U_{q}({\it sl}_{3})$
is completely the same as for any $ADE$ type algebra, so the construction
can be straightforwardly generalized to an arbitrary simply laced algebra.

\section {Construction of $U_{q}({\it sl}_{3})$}

The energy-momentum tensor corresponding to (\ref{action}) is given by:
\be
T(z)=-\frac{1}{2}(\partial\varphi(z),
\partial\varphi(z))+i\alpha_{\so}(\rho,\partial^{2}\varphi(z)),
\ee
where $( , )$ stands for the inner product, so
 $\varphi^{i}\varphi^{i}=(\varphi,\varphi)$.
$T(z)$ generates Virasoro algebra with central
 charge $c=n-12\alpha_{\so}^{2}(\rho,\rho)$.
Consider the vertex operators
$\int dz\,V_{\alpha}^{\spm}(\varphi)=$
\\
$\int dz\,:e^{i\alpha_{\spm}(\alpha,\varphi)}:$,
integrated over suitable contours.
Here $:\ldots:$ denotes a normal ordering prescription and  $\alpha_{\pm}$
are some constants obtained
from the following condition on the conformal dimension $\Delta$ of the vertex:
\be
\Delta(V_{\alpha}^{\spm})=\frac{1}{2}\,\,\alpha_{\spm}^{2}(\alpha,\alpha)
-\alpha_{\so}\alpha_{\spm}(\rho, \alpha)=1,
\ee
\be
\alpha_{\spm}=\frac{\alpha_{\so}}{2}\pm\sqrt{\left(\frac{\alpha_{\so}}{2}
\right)^{2}+1}.
\ee

In this formula we have assumed that $\alpha$ and $\beta$ are the simple
roots of $ {\it sl}_{3}$
and $\rho^{i}=\frac{1}{2}\sum_{+}\alpha^{i}$ is the Weyl vector, i. e.
$(\alpha,\alpha)=2$, $(\alpha,\beta)=-1$ and $(\rho,\alpha)=(\rho,\beta)=1$.
Consider the algebra with two generators:
\be
E_{\alpha}=\int dz :e^{i\alpha_{\spp}(\alpha,\varphi)}:,\,\,\,\,
E_{\beta}=\int dz :e^{i\alpha_{\spp}(\beta,\varphi)}:,
\ee
where the integration contour in the complex plane $z$
goes counterclockwise from the
point $z=1$ to the same point $z=1$ encircling the point $z=0$.
For the product of $n$ vertex operators we obtain the following expression:
\be
\label{int}
[\![ V_{\alpha_{1}}\ldots V_{\alpha_{n}}]\!]:=
\int_{\Gamma_{n}}dz_{1}\ldots
dz_{n}\,:e^{i\alpha_{\spp}(\alpha_{1},\,\varphi)}(z_{1}):\ldots :
e^{i\alpha_{\spp}(\alpha_{n},\,\varphi)}(z_{n}):\\
=\int_{\Gamma_{n}}dz_{1}\ldots dz_{n}\,\prod_{1\leq k<l\leq n}(z_{k}-z_{l})
^{\alpha_{\spp}^{2}(\alpha_{k},\,\alpha_{l})}
:e^{i\alpha_{\spp}(\alpha_{1},\,\varphi)}(z_{1})\ldots
e^{i\alpha_{\spp}(\alpha_{n},\,\varphi)}(z_{n}):,\\
\ee
where we use the Wick rule. The integration contour $\Gamma_{n}$ (fig.1.)
is chosen according to \cite{Felder:1989im}.
Namely, it consists of $n$ nested circles of unit radius
 oriented counterclockwise
from the point $z=1$ to $z=1$ around zero. The integral (\ref{int}) defined in this way contains
singularities, when $z_{k}=z_{l}$. We regularize it by the point-splitting
prescription with the parameter $\varepsilon$; the removal of the regularization takes
place in the limit $\varepsilon\rightarrow0$.

\begin{picture}(300,90)
\put(135,7){  $\scriptstyle{  Fig.1.\,Integration\,\,\,
 contour\,\,\, \Gamma_{n}   }$ .  }
\put(203,50){\ellipse{28}{25}}
\put(197,50){\circle{40}}
\put(211,50){\ellipse{12}{10}}
\multiput(195,50)(3,0){3}{\circle*{1}}
\put(170,42){$\scriptstyle{z_{1}}$}
\put(200,42){$\scriptstyle{z_{n}}$}
\put(182,42){$\scriptstyle{z_{2}}$}
\put(218,47){$\scriptstyle{1}$}
\end{picture}

The generator corresponding to the root
$\alpha+\beta$ is given by the $q$-deformed commutator:
\be
\label{third}
E_{\alpha+\beta}=-[E_{\alpha},E_{\beta}]_{q^{\sqm}}:=-[\![V_{\alpha}V_{\beta}
-q^{\sqm}V_{\beta}V_{\alpha}]\!]
\,\,\,\,\,;\,\,\,\,
q=q_{\spp}=e^{i\pi\alpha^{2}_{+}}.
\ee
This generator has a nonlocal form (see sec. 3). All other $q$-commutators
(of $E_{\alpha},E_{\beta}$ with $E_{\alpha+\beta}$)
are equal to zero, which is the
statement of Serre relation for the quantum group $U_{q}({\it sl}_{3})$:
\be
\label{serre}
[E_{\alpha},[E_{\alpha},E_{\beta}]_{q^{\sqm}}]_{q}=
[\![V_{\alpha}V_{\alpha}V_{\beta}
-(q+q^{\sqm})V_{\alpha}V_{\beta}V_{\alpha}+V_{\beta}V_{\alpha}V_{\alpha}]\!]=0
.
\ee
The proof is presented in the next section and is based on the regularization
of the expression (\ref{serre}) by the
point splitting prescription. The main observation in the
course of computation of (\ref{serre}) is that it is satisfied
in all orders in $\frac{1}{\varepsilon}$.

The formulas (\ref{third},\ref{serre})
define the commutation relations for the upper-triangular part of the quantum
group $U_{q}({\it sl}_{3})$ with the deformation parameter $q_{\spp}$.

There is the "classical" limit of this construction
$\alpha_{\so} \rightarrow 0$ ($q \rightarrow -1$),
when the action (\ref{action}) contains only the kinetic term.
This limit does not coincide with
the limit $q \rightarrow 1$, when the quantum group reduces to a Lie algebra.
Taking this limit we obtain the so-called Frenkel-Kac-Segal (FKS)
construction \cite{Frenkel:1989im,Goddard} in the bosonic string theory.
The q-commutation relation (\ref{third}) goes to
\be
E_{\alpha}E_{\beta}
-(-1)^{(\alpha,\beta)}E_{\beta}E_{\alpha},
\ee
and the factor $(-1)^{(\alpha,\beta)}$ is removed by
appropriate cocycles (see \cite{Goddard} for
details).

\section{Serre relation for $U_{q}({\it sl}_{3})$}

In this section we proof the Serre relation (\ref{serre})

Lets introduce the following notation for the integral ({\ref{int}}) with the ordering of the variables:
\be
\label{ordint}
I_{\alpha_{1}\ldots\alpha_{n}}=\int_{\arg z_{k}<\arg z_{k+1}-\varepsilon}
dz_{1}\ldots
dz_{n}\,:e^{i\alpha_{\spp}(\alpha_{1},\,\varphi)}(z_{1}):\ldots
 :e^{i\alpha_{\spp}
(\alpha_{n},\,\varphi)}(z_{n}):,\\
\ee
where $1\leq k \leq n-1$. We can construct the third generator
 explicitly, using
$q=e^{i\pi\alpha^{2}_{+}}$:
\be
\label{tri}
E_{\alpha+\beta}=-[\![V_{\alpha}V_{\beta}-q^{\sqm}V_{\beta}V_{\alpha}]\!]\\
=-\int_{\Gamma_{2}}dz_{1}\,dz_{2}\,(z_{1}-z_{2})^{-\,\alpha_{\spp}^{2}}
:e^{i\alpha_{\spp}(\alpha,\,\varphi)}(z_{1})
e^{i\alpha_{\spp}(\beta,\,\varphi)}(z_{2}):+
q^{\sqm}\cdot(\alpha\leftrightarrow\beta).
\ee

\begin{picture}(300,90)

\put(115,7)
{  $\scriptstyle  Fig.2.\,Contours\,\,\,of\,\,\,
integration\,\,\,in\,\,(\ref{tri})$.}
\put(103,50){\ellipse{28}{25}}
\put(97,50){\circle{40}}
\put(69,42){$\scriptstyle{z_{1}}$}
\put(81,42){$\scriptstyle{z_{2}}$}
\put(203,50){\ellipse{28}{25}}
\put(197,50){\circle{40}}
\put(169,42){$\scriptstyle{z_{2}}$}
\put(182,42){$\scriptstyle{z_{1}}$}
\put(303,50){\ellipse{33}{30}}
\put(278,42){$\scriptstyle{z_{2}}$}

%\put(319.3,50){\circle*{1}}
\put(300,64.5){\circle*{1}}
\spline(305.3,67.5)(304.7,68.95)(300,71)(295.05,68.95)(293,64)
(295.05,59.05)(300,57)
(304,59.05)(305.3,60.5)
\put(307,69){$\scriptstyle z_{1}$}
\qbezier(319,50)(319,65.5)(305.3,67.5)
\qbezier(319,50)(315,61)(305.3,60.5)

\put(140,50){\line(1,0){8}}
\put(240,50){=}
\end{picture}

Change of the variables $z_{1} \leftrightarrow z_{2}$
 in the second integral
produces compensating multiple factor $q$ and replaces
 the integration
contours as shown above on fig.2.
One could integrate separately around the $\varepsilon$-circle
($z_{1}-z_{2}=\varepsilon e^{i\psi}$) and rewrite the remaining
part in the form ({\ref{ordint}}):
\be
E_{\alpha+\beta}=-\varepsilon^{1-\alpha_{\spp}^{\skv}}\int id\psi\,dz_{2}\,
e^{i(1-\alpha_{\spp}^{2})\psi}:e^{i\alpha_{\spp}(\alpha,\,\varphi)}
(z_{2}+\varepsilon e^{i\psi})e^{i\alpha_{\spp}(\beta,\,\varphi)}
(z_{2}):-(1-q^{\smkv})I_{\alpha\beta}.
\ee
Here the factor $(1-q^{\smkv})$ appears due to $(z_{1}-z_{2})^
{-\,\alpha_{\spp}^{2}}$
in the integrand.
Note that in the classical limit $\alpha_{\so}\rightarrow0$
(or, equivalently $q\rightarrow-1$)
the coefficient at the nonlocal term $I_{\alpha\beta}$ vanishes
and we get the "classical" definition of the third generator
\cite{Frenkel:1989im,Goddard}: $E_{\alpha+\beta}=$\\
$\int dz
:e^{i(\alpha+\beta,\varphi)}$:.

The crucial point in our consideration is the Serre
identity (\ref{serre})
for $U_{q}({\it sl}_{3})$
quantum group. Let us rewrite it in the following way:
\be
\label{serre1}
[\![V_{\alpha}V_{\alpha}V_{\beta}-q^{\sqm}V_{\alpha}
V_{\beta}V_{\alpha}]\!]
-q[\![V_{\alpha}V_{\beta}V_{\alpha}-q^{\sqm}V_{\beta}
V_{\alpha}V_{\alpha}]\!]\\
=\int_{C_{1}}dz_{1}\,dz_{2}\,dz_{3}(z_{1}-z_{2})^
{2\,\alpha_{\spp}^{2}}
(z_{1}-z_{3})^{-\,\alpha_{\spp}^{2}}
(z_{2}-z_{3})^{-\,\alpha_{\spp}^{2}}
:e^{i\alpha_{\spp}(\alpha,\,\varphi)}(z_{1})\\
e^{i\alpha_{\spp}(\alpha,\,\varphi)}(z_{2})
e^{i\alpha_{\spp}(\beta,\,\varphi)}(z_{3}):
-q\int_{C_{2}^{'}}dz_{1}\,dz_{2}\,dz_{3}
(z_{1}-z_{2})^{-\,\alpha_{\spp}^{2}}
(z_{1}-z_{3})^{2\,\alpha_{\spp}^{2}}\\(z_{2}-z_{3})
^{-\,\alpha_{\spp}^{2}}
:e^{i\alpha_{\spp}(\alpha,\,\varphi)}(z_{1})
e^{i\alpha_{\spp}(\beta,\,\varphi)}(z_{2})e^{i\alpha
_{\spp}(\alpha,\,\varphi)}(z_{3}):.
\ee
Then changing the variables $z_{1} \rightarrow z_{2}, z_{2}
\rightarrow z_{3}, z_{3} \rightarrow
z_{1}$ in the second integral (after that contour $C_{2}^{'}$
is changing on
$C_{2}$ and multiple $q^{\sqm}$ appears)
and integrating over $\varepsilon$-circles: $z_{1}-z_{3}=
\varepsilon e^{i\theta}$,
$z_{2}-z_{3}=\varepsilon e^{i\psi}$
we obtain (see fig.3):
\be
\label{vych}
\varepsilon^{2}
\int id\theta\,id\psi\,dz_{3}\,
(e^{i\theta}-e^{i\psi})^{2\alpha_{\spp}^{\skv}}
e^{i(1-\alpha_{\spp}^{\skv})(\theta+\psi)}
:e^{i\alpha_{\spp}(\alpha\,\varphi)}(z_{3}+\varepsilon e^{i\theta})
e^{i\alpha_{\spp}(\alpha\,\varphi)}(z_{3}+\varepsilon e^{i\psi})\\
e^{i\alpha_{\spp}(\beta\,\varphi)}(z_{3}):+
\varepsilon^{1-\alpha_{\spp}^{\skv}}\int id\psi\,dz_{1}\,dz_{3}\,
(z_{1}-z_{3}-\varepsilon e^{i\psi})^{2\alpha_{\spp}^{\skv}}
(z_{1}-z_{3})^{-\alpha_{\spp}^{\skv}}
e^{i(1-\alpha_{\spp}^{\skv})\psi}\\
:e^{i\alpha_{\spp}(\alpha\,\varphi)}(z_{1})
e^{i\alpha_{\spp}(\alpha\,\varphi)}(z_{3}+\varepsilon e^{i\psi})
e^{i\alpha_{\spp}(\beta\,\varphi)}(z_{3}):+\\
\varepsilon^{1-\alpha_{\spp}^{\skv}}\int id\theta\,dz_{2}\,dz_{3}\,
(z_{3}-z_{2}+\varepsilon e^{i\theta})^{2\alpha_{\spp}^{\skv}}
(z_{2}-z_{3})^{-\alpha_{\spp}^{\skv}}
e^{i(1-\alpha_{\spp}^{\skv})\theta}\\
:e^{i\alpha_{\spp}(\alpha\,\varphi)}(z_{2})
e^{i\alpha_{\spp}(\alpha\,\varphi)}(z_{3}+\varepsilon e^{i\theta})
e^{i\alpha_{\spp}(\beta\,\varphi)}(z_{3}):+\\
\int_{C_{\varepsilon}}dz_{1}\,dz_{2}\,dz_{3}(z_{1}-z_{2})^{2\
,\alpha_{\spp}^{2}}
(z_{1}-z_{3})^{-\,\alpha_{\spp}^{2}}
(z_{2}-z_{3})^{-\,\alpha_{\spp}^{2}}\\
:e^{i\alpha_{\spp}(\alpha,\,\varphi)}(z_{1})
e^{i\alpha_{\spp}(\alpha,\,\varphi)}(z_{2})e^{i\alpha_{\spp}
(\beta,\,\varphi)}(z_{3}):.
\ee

\begin{picture}(300,95)
\put(40,7)
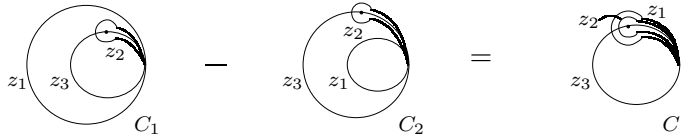
{  $\scriptstyle  Fig.3.\,\,Contours\,\,\,of\,\,\,integration\,\,\,in\,\,
(\ref{serre1},
\ref{vych})\\(\,all\,\,\,countours\,\,\,taken\,\,\,counterclockwise\,)$.}
\put(103,50){\ellipse{28}{25}}
\put(95,50){\circle{45}}
\put(65,42){$\scriptstyle{z_{1}}$}
\put(81,42){$\scriptstyle{z_{3}}$}
\put(113,25){$\scriptstyle{C_{1}}$}

%\put(216.7,50){\circle*{1}}
\put(199,69.8){\circle*{1}}
\spline(202.86,70.84)(202.46,71.8)(201.83,72.63)(199,73.8)
(196.17,72.63)(195,69.8)
(196.17,66.47)(199,65.7)(201.83,66.47)(202.40,67.6)
\qbezier(202.86,70.84)(214.9,68.0)(216.7,50)
\qbezier(202.42,67.5)(212.7,65.65)(216.7,50)
\put(192,61){$\scriptstyle z_{2}$}

\put(300,64.5){\circle*{1}}
\spline(303.46,66.5)(302.83,67.33)(300,68.5)(297.17,67.33)
(296,64.5)(297.17,61.17)
(300,60.4)(302.83,61.17)(303.46,62.5)
\qbezier(303.46,66.5)(317,65.5)(319,52)
\qbezier(303.46,62.5)(315,63)(319,52)

\qbezier(297.17,67.33)(293,70)(289,67)
\put(281,65){$\scriptstyle z_{2}$}

\put(102.5,62.5){\circle*{1}}
\spline(106.7,64.0)(105.6,66)(103,67.5)(99.5,66)
(98,62.5)(99.5,59)(104,58.7)(106.1,60)
\qbezier(106.7,64.0)(114,62.5)(117,50.5)
\qbezier(106.1,60)(112,60)(117,50.5)
\put(102,53){$\scriptstyle z_{2}$}

\put(205.2,50){\ellipse{23}{20}}
\put(197,50){\circle{40}}
\put(169,42){$\scriptstyle{z_{3}}$}
\put(186.5,42){$\scriptstyle{z_{1}}$}
\put(213,25){$\scriptstyle{C_{2}}$}

%\put(319,50){\circle*{1}}

\spline(305.3,67.5)(304.7,68.95)(300,71)(295.05,68.95)
(293,64)(295.05,59.05)(300,57)
(304,59.05)(305.3,60.5)
\put(307,69){$\scriptstyle z_{1}$}
\qbezier(319,50)(319,65.5)(305.3,67.5)
\qbezier(319,50)(315,61)(305.3,60.5)

\put(303,50){\ellipse{33}{30}}
\put(278,42){$\scriptstyle{z_{3}}$}
\put(313,25){$\scriptstyle{C}$}
\put(140,50){\line(1,0){8}}
\put(240,50){=}
\end{picture}

The first term in ({\ref{vych}}), which came from integration
around the two $\varepsilon$-circles, vanishes
in the limit $\varepsilon\rightarrow0$.
Divergent terms (second and third) cancel each other .
 The contour $C_{\varepsilon}$
in the fourth integral is the contour $C$ without
$\varepsilon$-circles. It
contains the integration over circle $|z_{1}-z_{2}|=
\varepsilon$, which is vanishing
in
the limit $\varepsilon\rightarrow0$. The remaining
integral might be rewritten in the
basis of $I_{\alpha_{1}\ldots\alpha_{n}}$ integrals
(\ref{ordint}):
\be
I_{\alpha\alpha\beta}+q^{\skv}q^{\smf}I_{\alpha\alpha\beta}-
q^{\skv}q^{\smf}I_{\alpha\alpha\beta}-q^{\smkv}q^{\skv}
I_{\alpha\alpha\beta}=0.
\ee
This completes the proof of the Serre identity.

\section {Quantum group $U_{q}( \widehat{\it sl}_{2})$ }

In order to generalize the construction above to the affine case
one should introduce two additional light-cone fields $\varphi_{+}$
and $\varphi_{-}$ and deform the action (\ref{action}) in the following way
\be
\label{bction}
S=\int d^{2}z\,(\partial\varphi^{i}\partial\varphi^{i}+
\partial\varphi_{+}\partial\varphi_{-}+i\alpha_{\so}R(\rho^{i}\varphi^{i}+
\rho_{+}\varphi_{-}+\rho_{-}\varphi_{+})).
\ee
Let $\beta_{\so}$ and $\beta_{\sqp}$ be the
simple roots of the affine Lie algebra $ \widehat{\it sl}_{2}$
with the Cartan matrix
\be
a_{ij}=\left(\begin{array}{rr}
2&-2\\-2&2
\end{array}\right).
\ee
Vertex operators of the conformal dimension 1 corresponding
to the simple roots are given by
\be
\label{vert}
E_{\so}=\int dz\,e^{i\alpha_{\spp}((\beta_{\so},\,\varphi)+
\varphi_{+})},\,\,\,\,
E_{\sqp}=\int dz\,e^{i\alpha_{\spp}(\beta_{\sqp},\,\varphi)}.
\ee
Let us write down those commutation relations
of $U_{q}( \widehat{\it sl}_{2})$, which contain only simple
root generators:
\be
\label{com1}
e_{\delta}=[E_{\so},E_{\sqp}]_{q^{\smkv}},
\ee
\be
\label{com2}
e_{n}=A(n,q)[E_{\sqp},[E_{\so},[E_{\sqp},[E_{\so},...
[E_{\sqp},[E_{\so},E_{\sqp}]_{q^{\smkv}}]...
]_{q^{\skv}}]]_{q^{\skv}}],
\ee
\be
\label{com3}
e_{n\delta}=B(n,q)[E_{\so},e_{n-1}]_{q^{\skv}},
\ee
where $A(n,q)$ and $B(n,q)$ are some normalizing constants. The
generator $e_{n}$ corresponds to the root $\beta_{\sqp}+n\delta$.
The Serre relations read:
\be
[e_{n+1},e_{m}]_{q^{\skv}}+[e_{m+1},e_{n}]_{q^{\skv}}=0,
\ee
\be
\label{sss}
[E_{\sqp},[E_{\sqp},[E_{\sqp},E_{\so}]_{q^{\skv}}]]_{q^{\smkv}}=
0,\,\,\,\,\,\,\,(m=n=0).
\ee
The proof of the Serre relation (\ref{sss}) is similar to the case of $U_{q}({\it sl}_{3})$
and is presented in the Appendix. We provide also the formulas for the generators (\ref{com2}, \ref{com3}).

In the classical limit ( $\alpha_{\so} \rightarrow 0$ )
the generators (\ref{vert})
reduce to:
\be
E_{\so}=\int dz\,e^{i((\beta_{\so},\,\varphi)+\varphi_{+})},\,\,\,\,
E_{\sqp}=\int dz\,e^{i(\beta_{\sqp},\,\varphi)},
\ee
which coincides with the FKS construction of $ \widehat{\it sl}_{2}$
\cite{Frenkel:1989im,Goddard} .

\section {Concluding remarks}

We have described the vertex operator construction
of the quantum group $U_{q_{+}}(n_{+})$,
where $n_{+}$ is the Borel subalgebra of some Lie algebra $g$.
By analogy one can construct $U_{q_{-}}(n_{-})$. It still
remains unclear how to combine these two parts into one object.

This type of the construction (using only scalars
$\varphi^{i}$) also can be applied to
the following underlying algebras: finite-dimensional $A, D, E$-series,
affine $A^{(1)}, D^{(1)}, E^{(1)}$ and, possibly, to some hyperbolic
algebras.

We have discussed the relationship between the vertex operator
realization of quantum groups and the FKS construction. Possible applications
to the quantization of conformal affine Toda theories will be studied elsewhere.

\vspace{0.2in}
{\large\bf
Acknowledgements
}
\vspace{0.2in}

The author is grateful to A. Gerasimov for numerous valuable
discussions, guidance and encouragement.
I would like to thank A. Alexandrov, V. Dolgushev, A. Dymarsky, S. Kharchev, A. Konechny, V. Poberezhny and
A. Zotov for useful discussions and E. Akhmedov, A. Morozov and E. Suslova for stimulating suggestions and advices.
The work was partly supported by the Russian President's grant 00-15-99296,
RFBR grant 07-02-00878, by the grant for support of scientific schools NSh-8004.2006.2, by the INTAS grant 00-334,
and by Federal agency for atomic energy of Russia.

\vspace{0.3in}

{\Large\bf Appendix
}
\vspace{0.3in}

In this appendix we present the proof of the Serre relations for
$U_{q}( \widehat{\it sl}_{2})$.
We also provide some formulas for the generators
corresponding to the positive roots of $U_{q}( \widehat{\it sl}_{2})$.
First, let us compute the light-like root generator (\ref{com1})
(see appendix A for the details of calculation):
\be
e_{\delta}=[E_{\so},E_{\sqp}]_{q^{-\skv}}=
\varepsilon^{1-2\alpha_{\spp}^{\skv}}\int id\psi\,dz\,
e^{i(1-2\alpha_{\spp}^{2})\psi}:
e^{i\alpha_{\spp}((\beta{\so},\,\varphi)+\varphi_{+})}
(z+\varepsilon
e^{i\psi})\\
e^{i\alpha_{\spp}(\beta_{\sqp},\,\varphi)}(z):+(1-q^{\smf})
I_{01}.
\ee
In the classical limit ($\alpha{\so}\rightarrow 0$)
the second term vanish
and the first
integral reduces to $e_{\delta}=\int dz\,\partial\varphi
e^{\varphi_{+}}$.

The proof of Serre relation ({\ref{sss}}):
\be
\label{afser}
[E_{\sqp},[E_{\sqp},[E_{\sqp},E_{\so}]_{q^{\skv}}\,]\,]_{q^{\smkv}}=0,
\ee
reduces to the proof that the following integral is vanishing:
\be
\label{zero}
\int_{\Gamma} dz_{1}\,dz_{2}\,dz_{3}\,dz_{4}
(z_{1}-z_{2})^{2\,\alpha_{\spp}^{2}}(z_{1}-z_{3})^{2\,\alpha_{\spp}^{2}}
(z_{1}-z_{4})^{-2\,\alpha_{\spp}^{2}}(z_{2}-z_{3})^{2\,\alpha_{\spp}^{2}}
(z_{2}-z_{4})^{-2\,\alpha_{\spp}^{2}}\\
(z_{3}-z_{4})^{-2\,\alpha_{\spp}^{2}}
:e^{i\alpha_{\spp}(\beta_{\sqp},\,\varphi)}(z_{1})
e^{i\alpha_{\spp}(\beta{\sqp},\,\varphi)}(z_{2})e^{i\alpha_{\spp}
(\beta{\sqp},\,\varphi)}(z_{3})
e^{i\alpha_{\spp}((\beta{\so},\,\varphi)+\varphi_{+})}(z_{4}):=0.
\ee
The contour $\Gamma$ is constructed as follows.
 Each variable $z_{1},z_{2},z_{3}$ is
integrated from a point 1 to 1 encircling point $z_{4}$,
the variable $z_{4}$ is integrated afterwards over the unit circle.
The contours of integration
around $z_{1},z_{2},z_{3}$ are nested.

It is convenient to introduce the following notation.
Let $\delta_{j}$ ($1<j<3$) denote the contour of integration for the
variable $z_{j}$ that consists of an $\varepsilon$-circle
around $z_{4}$:
$z_{j}-z_{4}=\varepsilon e^{i\theta_{j}}$. Also denote by $\gamma_{j}$
 a contour
of integration for variable $z_{j}$ that consists of two parts:
first, we
integrate counterclockwise around the arc
$0<\mathrm{arg} z_{j}<\mathrm{arg} z_{4}-\varepsilon$, $|z_{j}|>|z_{4}|$ and
secondly, clockwise around the arc $0<\mathrm{arg} z_{j}<\mathrm{arg} z_{4}-\varepsilon$,
$|z_{j}|<|z_{4}|$.

According to this notations the integration contour $\Gamma$ in
(\ref{zero}) is separated into 8 pieces:

1. Integration over $\delta_{1}$, $\delta_{2}$, $\delta_{3}$ and $z_{4}$.
This integral is
proportional to $\varepsilon^{3}$ and vanishes in the limit
$\varepsilon \rightarrow 0$.

2. $\delta_{1}$, $\gamma_{2}$, $\gamma_{3}$ and $z_{4}$,

3. $\gamma_{1}$, $\delta_{2}$, $\gamma_{3}$ and $z_{4}$,

4. $\gamma_{1}$, $\gamma_{2}$, $\delta_{3}$ and $z_{4}$,\\
Each of these integrals is proportional to the following integral:
\be
\varepsilon^{2(1-\alpha_{\spp}^{\skv})}\int id\chi\,dx\,dy\,dz\,
(x-y)^{2\alpha_{\spp}^{\skv}}(x-z-\varepsilon e^{i\chi})
^{2\alpha_{\spp}^{\skv}}
(x-z)^{-2\alpha_{\spp}^{\skv}}(y-z-\varepsilon e^{i\chi})
^{2\alpha_{\spp}^{\skv}}\\
(y-z)^{-2\alpha_{\spp}^{\skv}}
e^{i(1-2\alpha_{\spp}^{\skv})\chi}
:e^{i\alpha_{\spp}(\beta{\sqp},\varphi)}(x)e^{i\alpha_{\spp}
(\beta{\sqp},\varphi)}(y)
e^{i\alpha_{\spp}(\beta{\sqp},\varphi)}(z+\varepsilon e^{i\chi})
e^{i\alpha_{\spp}((\beta{\so},\,\varphi)+\varphi_{+})}(z):.
\ee
Here the integration over $x$, $y$, $z$ is ordered, as in (\ref{ordint}).
 The sum of
the proportionality coefficients is equal to zero. Similarly the
integrals over the:

5. $\delta_{1}$, $\delta_{2}$, $\gamma_{3}$ and $z_{4}$,

6. $\delta_{1}$, $\gamma_{2}$, $\delta_{3}$ and $z_{4}$,

7. $\gamma_{1}$, $\delta_{2}$, $\delta_{3}$ and $z_{4}$\\
cancel each other.

8. $\gamma_{1}$, $\gamma_{2}$, $\gamma_{3}$ and $z_{4}$.\\
The integration over splitting points $z_{1}$,$z_{2}$
and $z_{3}$ does not contribute
in the limit $\varepsilon \rightarrow 0$.
The remaining integral can be rewritten in the basis
of $I_{\alpha_{1}\ldots\alpha_{n}}$ (\ref{ordint}) and also vanishes.

Let us perform here the computations of some generators in the affine case.
They are based on the following formulas:
\be
\label{f1}
[E_{\so}, I_{\epsilon_{0} \epsilon_{1}...\epsilon_{n-1} \epsilon_{n}}]=
(1-q^{\smf}) \sum_{i=0}^{n}[a_{i}+1]_{q^{\skv}}I_{\epsilon_{0}
\epsilon_{1}...\epsilon_{i-1} \so \epsilon_{i}
...\epsilon_{n-1} \epsilon_{n}},
\ee
\be
\label{f2}
[E_{\sqp}, I_{\epsilon_{0} \epsilon_{1}...
\epsilon_{n-1} \epsilon_{n}}]=
(q^{\skv}-q^{\smkv}) \sum_{i=0}^{n}[b_{i}]
_{q^{\skv}}I_{\epsilon_{0} \epsilon_{1}
...\epsilon_{i-1} \sqp \epsilon_{i}
...\epsilon_{n-1} \epsilon_{n}},
\ee
where:
\be
a_{i}=\frac{1}{2}\sum_{k=0}^{i-1}(\beta_{\epsilon_{k}}
,\beta_{0}),\,\,\,\,\,\,\,\,
b_{i}=\frac{1}{2}\sum_{k=0}^{i-1}(\beta_{\epsilon_{k}}
,\beta_{1}).
\ee
In these formulas $I_{\epsilon_{0} \epsilon_{1}...
\epsilon_{n-1} \epsilon_{n}}$
stands for $I_{\beta_{\epsilon_{0}} \beta{\epsilon_{1}}
...\beta{\epsilon_{n-1}}
\beta_{\epsilon_{n}}}$ ($\epsilon_{i}=0$ or $1$)
 (see (\ref{ordint})).

The expressions for $e_{n}$ and $e_{n \delta}$ can
be written in the same
manner as (\ref{int}) and (\ref{vych}).
We can analytically continue the integrands in these
expressions (regarding
$\alpha_{\spp}^{2}$ as a complex parameter, see e. g.
 \cite{fateev1}) from the region, where the integrand
is well-defined, so that $\varepsilon$-terms are not significant.
Doing this and using the formulas above the generators
 (\ref{com2}) can be written down in the basis of
$I_{\epsilon_{0} \epsilon_{1}...\epsilon_{n-1}
\epsilon_{n}}$-integrals:
\be
\label{f3}
e_{n}=A(n,q)
q^{\skv n}(1-q^{\smf})^{\skv n} \sum_{\epsilon_{1}+...
+\epsilon_{2n-1}=n} c_{\so \epsilon_{1}
...\epsilon_{2n-1}\sqp} I_{\so \epsilon_{1}
...\epsilon_{2n-1}\sqp},
\ee
where summation goes over all nontrivial permutations
of $\epsilon_{1}, ...
,\epsilon_{2n-1}$. The coefficients
$ c_{\epsilon_{\so} \epsilon_{1}...\epsilon_{2n-1} \epsilon_{2n}}$
are given by the following expression:
\be
\label{f4}
c_{\epsilon_{\so} \epsilon_{1}...\epsilon_{2n-1} \epsilon_{2n}}=\\
\sum_{j_{1}(\epsilon_{j_{1}}=1)} \sum_{j_{2} \neq j_{1}
(\epsilon_{j_{2}}=0)}...
\sum_{j_{2n-1} \neq j_{2n-2} \neq...\neq j_{1} (\epsilon_{j_{2n-1}}=1)}
[b_{j_{1}}]_{q^{\skv}}[a'_{j_{2}}+1]_{q^{\skv}} ...
[a'_{j_{2n-2}}+1]_{q^{\skv}}[b'_{j_{2n-1}}]_{q^{\skv}},
\ee
where we use the notations:
\be
a'_{j_{2l}}=\frac{1}{2}\sum_{k=0,\phantom{1} k\neq
j_1,...,j_{2l-1}}^{j_{2l}-1}
(\beta_{\epsilon_{k}},\beta_{0})\,\,\,\,\,\,\,\,\,\,\,\,\,\,\,\,\
b'_{j_{2l-1}}=\frac{1}{2}\sum_{k=0,\phantom{1} k\neq
j_1,...,j_{2l-2}}^{j_{2l-1}-1}(\beta_{\epsilon_{k}},\beta_{1}).
\ee

The symbol $\sum_{j_{k}(\epsilon_{j_{k}}=1)}$
 denotes summation over all
$j_{k}$ ($1\leq j_{k}\leq 2n-1$) such that $\epsilon_{j_{k}}=1$. The
formulas for $e_{n \delta}$ can be written down, using
(\ref{com3},\ref{f1}).
For example, using (\ref{f1},\ref{f2},\ref{f3},\ref{f4})
one can obtain:
\be
e_{\delta}=(1-q^{\smf})I_{01},
\ee

\be
e_{\sqp}=[E_{\sqp},[E_{\so},E_{\sqp}]_{q^{\smkv}}]=-q^{\skv}
(1-q^{\smf})^{2}I_{011},
\ee

\be
e_{2\delta}=[E_{\so},[E_{\sqp},[E_{\so},E_{\sqp}]_{q^{\smkv}}]]_
{q^{\skv}}=-q^{\skv}(1-q^{\smf})^{3}
(I_{0101}+(1+[2]_{q^{\skv}})I_{0011}),
\ee

\be
e_{2}=[E_{\sqp},[E_{\so},[E_{\sqp},[E_{\so},E_{\sqp}]_{q^{\smkv}}]]_
{q^{\skv}}]=\\
q^{\spf}(1-q^{\smf})^{\spf}
(I_{01101}+(2+[2]_{q^{\skv}})I_{01011}+(1+[2]_{q^{\skv}})^{\skv}
I_{00111}),
\ee

\be
e_{3\delta}=[E_{\so},[E_{\sqp},[E_{\so},[E_{\sqp},[E_{\so},E_{\sqp}]
_{q^{\smkv}}]]_{q^{\skv}}]]_{q^{\skv}}=
q^{\spf}(1-q^{\smf})^{5}
((1+[2]_{q^{\skv}})(2+[2]_{q^{\skv}})I_{001101}\\+(3+[2]_{q^{\skv}})I_{010101}+
I_{011001}+
(1+[2]_{q^{\skv}})^{\skv}(2+[2]_{q^{\skv}})I_{010011}+\\
(1+[2]_{q^{\skv}})(2+[2]_{q^{\skv}}+
 [2]_{q^{\skv}}(2+[2]_{q^{\skv}}))I_{001011}+(1+[2]_{q^{\skv}})^{\skv}
(1+[2]_{q^{\skv}}+[3]_{q^{\skv}})I_{000111}),
\ee
up to constants $A(n,q)$, $B(n,q)$.

\end{document}